\documentclass[runningheads]{llncs}
\usepackage{graphicx}
\usepackage{amsmath,mathtools}
\usepackage{graphicx}
\usepackage{array}
\usepackage{verbatim}
\usepackage{multirow}
\usepackage{multicol}
\usepackage[table]{xcolor}
\usepackage[T1]{fontenc}

\newcolumntype{L}[1]{>{\raggedright\let\newline\\\arraybackslash\hspace{0pt}}m{#1}}
\newcolumntype{C}[1]{>{\centering\let\newline\\\arraybackslash\hspace{0pt}}m{#1}}
\newcolumntype{R}[1]{>{\raggedleft\let\newline\\\arraybackslash\hspace{0pt}}m{#1}}

\begin{document}
\title{Radiomics Boosts Deep Learning Model for IPMN Classification}

%

\author{Lanhong Yao\inst{1} \and
Zheyuan Zhang\inst{1} \and
Ugur Demir\inst{1} \and
Elif Keles\inst{1} \and
Camila Vendrami\inst{1} \and
Emil Agarunov \inst{2} \and
Candice Bolan \inst{3} \and
Ivo Schoots\inst{4} \and
Marc Bruno \inst{4} \and
Rajesh Keswani \inst{1} \and
Frank Miller\inst{1} \and
Tamas Gonda \inst{2} \and
Cemal Yazici\inst{5} \and
Temel Tirkes \inst{6} \and
Michael Wallace \inst{7} \and
Concetto Spampinato \inst{8} \and
Ulas Bagci\inst{1} \thanks{This project is supported by the NIH funding: NIH/NCI R01-CA246704 and NIH/NIDDK U01-DK127384-02S1.}}
\authorrunning{L. Yao et al.}
%
%
%

\institute{Department of Radiology, Northwestern University, Chicago IL 60611, USA \and 
NYU Langone Health, New York, NY 10016 \and
Mayo Clinic, Rochester, MN 55905\and
Erasmus Medical Center, 3015 GD Rotterdam, Netherlands \and
University of Illinois Chicago, Chicago, IL 60607\and
Indiana University–Purdue University Indianapolis, Indianapolis, IN 46202\and
Sheikh Shakhbout Medical City, 11001, Abu Dhabi, United Arab Emirates\and
University of Catania, 95124 Catania CT, Italy}
\maketitle              
\begin{abstract}
Intraductal Papillary Mucinous Neoplasm (IPMN) cysts are pre-malignant pancreas lesions, and they can progress into pancreatic cancer. Therefore, detecting and stratifying their risk level is of ultimate importance for effective treatment planning and disease control. However, this is a highly challenging task because of the diverse and irregular shape, texture, and size of the IPMN cysts as well as the pancreas. In this study,  we propose a novel computer-aided diagnosis pipeline for IPMN risk classification from multi-contrast MRI scans. Our proposed analysis framework includes an efficient volumetric self-adapting segmentation strategy for pancreas delineation, followed by a newly designed deep learning-based classification scheme with a radiomics-based predictive approach. We test our proposed decision-fusion model in multi-center data sets of 246 multi-contrast MRI scans and obtain superior performance to the state of the art (SOTA) in this field. Our ablation studies demonstrate the significance of both radiomics and deep learning modules for achieving the new SOTA performance compared to international guidelines and published studies (81.9\% vs 61.3\% in accuracy). Our findings have important implications for clinical decision-making. In a series of rigorous experiments on multi-center data sets (246 MRI scans from five centers), we achieved unprecedented performance (81.9\% accuracy). The code is available upon publication. 

\end{abstract}
\section{Introduction}

Pancreatic cancer is a deadly disease with a low 5-year survival rate, primarily because it is often diagnosed at a late stage~\cite{chen2023pancreatic,lalonde2019inn}. Early detection is crucial for improving survival rates and gaining a better understanding of tumor pathophysiology. Therefore, research on pancreatic cysts is significant, since some types, such as intraductal papillary mucinous neoplasms (IPMN), can potentially develop into pancreatic cancer~\cite{luo2019characteristics}. Hence, the diagnosis of IPMN cysts and the prediction of their likelihood of transforming into pancreatic cancer are essential for early detection and disease management. Our study is aligned with this objective and aims to contribute to aiding the early detection of pancreatic cancer.

\begin{figure}
\centering
{\includegraphics[width=\textwidth]{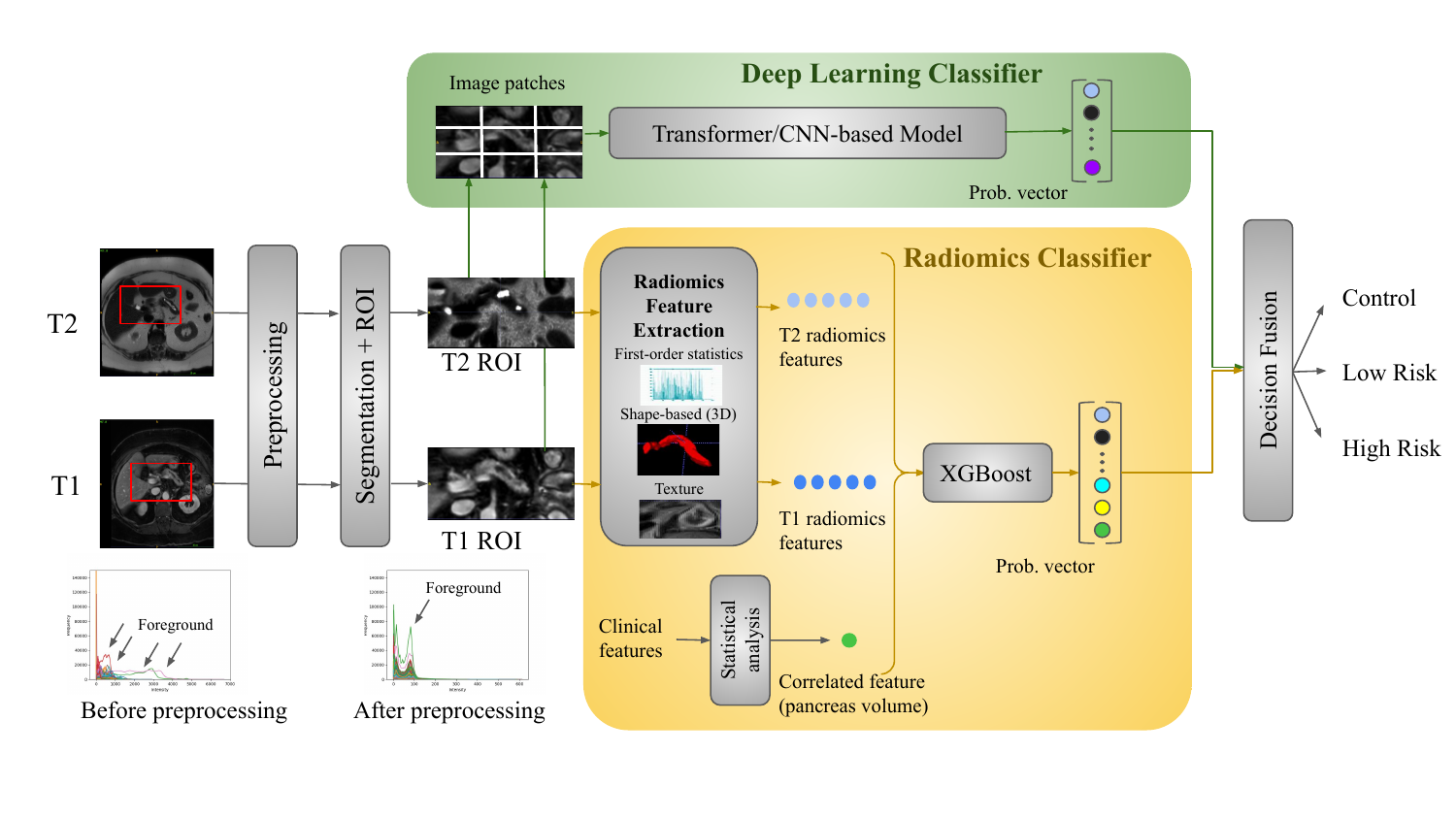}}
\caption{An overview of our proposed CAD system is shown. Multi-contrast MRI (T1 and T2) are preprocessed with inhomogeneity correction, denoising, and intensity standardization. Clean images are then used to segment the pancreas region. ROI enclosing the segmented pancreases are fed into a \textbf{deep learning classifier}. Clinical features selected through statistical analysis are fed into \textbf{radiomics classifier}. Decision vectors (probability) from both classifiers are combined via a weighted averaging-based decision fusion strategy for final IPMN cyst stratification.}
\label{overview}
\end{figure}

Diagnosis of IPMN involves a combination of imaging studies, laboratory tests, and sometimes biopsy. Imaging studies utilize CT, MRI, and EUS scans to visualize the pancreas and detect any cystic lesions. The size, location, shape, texture, and other characteristics of the lesions are used for radiographical evaluations. Current international guidelines (AGA, ACG, and IAP)~\cite{elta2018acg,lennon2015aga,marchegiani2018systematic} state that IPMNs should be classified into low-risk or high-risk based on their size, morphology, and presence of high-risk features such as main pancreatic duct involvement, mural nodules, or elevated cyst fluid CEA levels. While high-risk IPMNs should be considered for surgical resection, low-risk IPMNs may be managed with surveillance.

\subsubsection{Prior Art.} Radiographical identification of IPMNs is important but falls short of diagnostic accuracy, thus there is a need for improving the current standards for IPMN risk stratifications~\cite{kuwahara2019usefulness}. Several studies have demonstrated the potential of deep learning in IPMN diagnosis, including the use of convolutional neural networks (CNN)~\cite{sarfaraz}, inflated neural networks (INN)~\cite{lalonde2019inn}, and neural transformers~\cite{salanitri2022neural}. However, these studies analyzed MRIs from a single center with a small number of patients and did not include the pancreas segmentation step, only using directly cropped pancreas regions or whole images for classification. Still, these models showed promising results compared to the international guidelines. While deep learning techniques have shown the potential to improve the accuracy and efficiency of IPMN diagnosis and risk stratification~\cite{corral2019deep}, further research and validation are needed to determine the clinical utility and their potential impact on patient outcomes. Our work addresses the limitations of current deep learning models by designing a new computer-aided diagnosis (CAD) system, which includes a fully automated pipeline including (1) MRI cleaning with preprocessing, (2) segmentation of pancreas, (3) classification with decision fusion of deep learning and radiomics, (4) statistical analysis of clinical features and incorporation of the pancreas volume into clinical decision system, and (5) testing and validation of the whole system in the multi-center settings. Figure~1 shows an overview of the proposed novel CAD system. 

\subsubsection{Summary of Our Contributions.} To the best of our knowledge, this is the first study having a fully automated pipeline for IPMN diagnosis and risk stratification, developed and evaluated on multi-center data. Our major contributions are as follows:
\begin{enumerate}
    \item We develop the first fully automated CAD system that utilizes a powerful combination of deep learning, radiomics, and clinical features - all integrated into a single decision support system via a weighted averaging-based decision fusion strategy. 
   \item Unlike existing IPMN CAD systems, which do not include MRI segmentation of the pancreas and require cumbersome manual annotations on the pancreas, we incorporate the volumetric self-adapting segmentation network (nnUNet) to effectively segment the pancreas from MRI scans with high accuracy and ease. 
    \item We present a simple yet convenient approach to fusing radiomics that can significantly improve the DL model's performance by up to ~20\%. This powerful tool allows us to deliver more accurate diagnoses for IPMN.
    \item  Through rigorous statistical analysis of 8 clinical features, we identify pancreas volume as a potential predictor of risk levels of IPMN cysts. By leveraging this vital information, we enhance the decision fusion mechanism to achieve exceptional overall model accuracy.
\end{enumerate}

\section{Materials and Methods}
\subsection{Dataset}
In compliance with ethical standards, our study is approved by the Institutional Review Board (IRB), and necessary privacy considerations are taken into account: all images are de-identified before usage. 
We obtain 246 MRI scans (both T1 and T2) from five centers: Mayo Clinic in Florida (MCF), Mayo Clinic in Arizona (MCA), Allegheny Health Network (AHN), Northwestern Memorial Hospital (NMH), and New York University Langone Hospital (NYU). All T1 and T2 images are registered and segmentation masks are generated using a fast and reliable segmentation network (see below for the segmentation section). Segmentations are examined by radiologists case by case to ensure their correctness. The ground truth labels of IPMN risk classifications are determined based on either biopsy exams or surveillance information with radiographical evaluation, and overall three balanced classes are considered for risk stratification experiments: healthy (70 cases), low-grade risk (85 cases), and high-grade risk (91 cases).

\subsection{Preprocessing}
MRI presents unique challenges, including intensity inhomogeneities, noise, non-standardization, and other artifacts. These challenges often arise from variations in acquisition parameters and hardware, even when using the same scanner and operators at different times of the day or with different patients. Therefore, preprocessing MRI scans across different acquisitions, scanners, and patient populations is necessary. In our study, we perform the following preprocessing steps on the images before feeding them into the segmentor and classifiers: Initially, images are reoriented in accordance with the RAS axes convention. Subsequent steps involve the application of bias correction and denoising methodologies, designed to mitigate artifacts and augment image fidelity. Further, we employ Nyul's method~\cite{nyul2000new} for intensity standardization, harmonizing the intensity values of each image with a designated reference distribution. Figure 1 illustrates the image histograms pre- and post-preprocessing, underscoring the efficacy of our standardization procedure. These preprocessing steps can help improve the robustness and reliability of deep learning models. 

\vspace{-2mm}
\subsection{Pancreas Segmentation}
\vspace{-1mm}
Pancreas volumetry is a prerequisite for the diagnosis and prognosis of several pancreatic diseases, requiring radiology scans to be segmented automatically as manual annotation is highly costly and inefficient. In this module of the CAD system, our aim is to develop a clinically stable and accurate deep learning-based segmentation algorithm for the pancreas from MRI scans in multi-center settings to prove its generalization efficacy. Among 246 scans, we randomly select 131 MRI images (T2) from multi-center settings: 61 cases from NMH, 15 cases from NYU, and 55 cases from MCF. Annotations are obtained from three centers' data. The segmentation masks are used for pancreas region of interest (ROI) boundary extraction in radiomics and deep learning-based classification rather than exact pixel analysis. We present a robust and accurate deep learning algorithm based on the 3D nnUNet architecture with SGD optimization \cite{isensee2021nnunet}. 

\subsection{Model Building for Risk Stratification}
\subsubsection{Radiomics Classifier}

Radiomics involves extracting a large number of quantitative features from medical images, offering valuable insights into disease characterization, notably in IPMN where shape and texture are vital to classifications. For this study, 107 distinct features are extracted from both T1 and T2 images, within the ROI  enclosing pancreas segmentation mask. They capture characteristics such as texture, shape, and intensity. To mitigate disparities across scales in the radiomics data, we employ the $ln(x+1)$ transformation and unit variance scaling. Further, analyze 8 clinical features using an OLS regression model to evaluate their predictive efficacy for IPMN risk: diabetes mellitus, pancreas volume, pancreas diagonal, volume over diagonal ratio, age, gender, BMI, and chronic pancreatitis. Through stepwise regression, we refine the model's focus to key features. T-tests reveal significant differences in pancreas volume across IPMN risk groups, consistent with prior medical knowledge. Notably, pancreas volume shows predictive efficacy for IPMN risk, leading to its inclusion as a vital clinical feature in our risk prediction model.

\subsubsection{Deep Learning Classifier}
We utilize one Transformer-based and four CNN-based architectures to compare and evaluate the IPMN risk assessment. Neural transformers~\cite{salanitri2022neural} is notably the first application of vision transformers (ViT) in pancreas risk stratification and has obtained promising results but with a limited size of data from a single center. DenseNet~\cite{huang2017densely}, ResNet18~\cite{he2016deep}, AlexNet~\cite{krizhevsky2017imagenet}, and MobileNet~\cite{sandler2018mobilenetv2} are all well-known CNN-based architectures that have been developed in recent years and have been shown to perform well on various computer vision tasks.  Herein, we benchmark these models to create baselines and compare/contrast their benefits and limitations. 
 
\subsubsection{Weighted averaging based decision fusion}
The sparse feature vectors learned by DL models pose a challenge for feature fusion with the radiomics model. To address this, we implement weighted averaging-based decision fusion, where we combine the weighted probabilities of the DL classifier (shown below in Eq. 1) and radiomics classifier: 
\begin{equation}
P_{c}=
\begin{cases}
P_{r},& \text{if } max(P_{r}) \geq t\\
k*P_{d} + (1-k)*P_{r},              & \text{otherwise}
\end{cases}
\end{equation}
where $k$ and $t$ are parameters to adjust the decision fusion of two models, and they are selected via grid search during cross-validation. \textit{P\textsubscript{d}} and \textit{P\textsubscript{r}} are probability vectors, predicted by the DL classifier and radiomics classifier, respectively. \textit{P\textsubscript{c}} refers to the combined probabilities, based on which we get the final fused predictions. For each case, \textit{P\textsubscript{c}} for three classes add up to one. This \textit{P\textsubscript{c}} is used in the cases when the maximum probability of the radiomics classifier is less than a threshold $t$, indicating the radiomics classifier is not confident in its decision and could use extra information.

\subsection{Training Details}


The dataset used in this study consists of 246 patients from five medical centers. Out of these, 49 cases are randomly selected for blind testing (i.e., independent test), and are unseen by any of the models. The remaining 197 cases are split into training and validation sets. Every set incorporates data across all participating medical centers. This consistent distribution ensures an unbiased evaluation environment. We employ the same evaluation procedure for all the models.

The deep learning (DL) models have been developed utilizing the PyTorch framework and executed on an NVIDIA RTX A6000 GPU. The training was conducted with a batch size of 16 across a maximum of 1500 epochs. The radiomics classifier employs XGBOOST with grid search to identify the best parameters. The optimal parameters are determined as follows: number of estimators=140 and maximum depth=4. 

\section{Results}
\begin{figure}[t]
    \centering
    \includegraphics[width=0.9\textwidth]{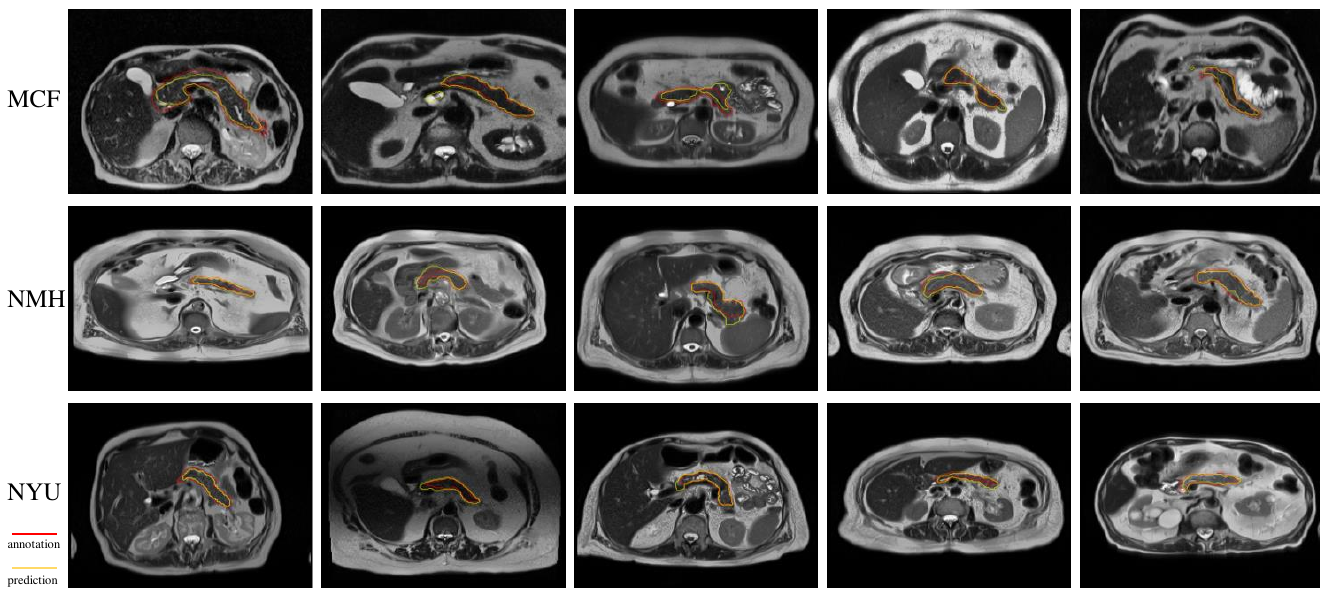}
    \caption{Qualitative pancreas segmentation visualization on multi-center data. Predicted segmentation maps (yellow line) are highly similar to the ground truth annotations (red line) in anatomical structure regardless of image variance.}
    \label{fig:segvisual}
\end{figure}
\subsection{Segmentation}
We employ standard 5-fold cross-validation for the training and use Dice score (higher is better), Hausdorff distance at 95\% (HD95, lower is better), Precision, and Recall for quantitative evaluations. Dice score for CT-based segmentation in the literature reaches a plateau value of 85\% but MRI segmentations hardly reach 50-60\% with a limited number of research papers \cite{zhang2022dynamic}. Herein, our segmentation results reach 70\% for multi-center data, showing a significant increase from the current standards. Fig. \ref{fig:segvisual} shows qualitative visualization of the predicted segmentation results compared with the reference standards provided by the radiologists, demonstrating highly accurate segmentations. 

We conduct a comprehensive quantitative evaluation. Table 1 shows the quantitative evaluation results on multi-domain, and particularly, we reach one average Dice of 70.11\% which has never been achieved before in the literature in multi-center settings.

\begin{table}
\centering
\caption{Multi-center pancreas  MRI segmentation performance comparison. The model reaches a sufficiently accurate segmentation with average Dice of 70.11.}
\begin{tabular}{ c | *{4}{ C{2.1cm}@{}} |c }
\hline
Data Center & Dice       & HD95(mm)     & Precision  & Recall  & Case   \\
\hline
MCF                     & 69.51±2.96 & 28.36±23.59 & 77.18±3.50 & 66.11±4.36 & 61\\
NMH                     & 66.02±1.84 & 14.91±9.95  & 63.39±4.30 & 71.19±6.23 & 15\\
NYU                    & 71.90±3.20 & 26.59±12.00  & 75.9±1.59  & 71.73±4.56 & 55 \\
\hline
   \rowcolor[gray]{0.9} Average                 & 70.11±2.96 & 26.08±18.19 & 75.06±2.98 & 69.05±4.70 & Sum:131\\
\hline
\end{tabular}
\end{table}

\vspace{-2mm}
\subsection{Classification}

\begin{table}[!b]
\centering
\caption{Quantitative comparison (\%) for the influence of combining radiomics with deep learning. We can observe that regardless of network structure, combining radiomics with deep learning can impressively leverage the IPMN classification performance. Similarly, combining the deep learning extracted features can also leverage the performance of radiomics for the IPMN classification.}
\begin{tabular}{c | *{4}{ C{1.1cm}@{}} |*{4}{ C{1.1cm}@{}} }
  \hline
  \hline
  Network & \multicolumn{4}{c|}{\textbf{w/o} Radiomics} &\multicolumn{4}{c}{\textbf{w/} Radiomics} \\
  \cline{0-8}
Metrics & ACC & AUC & PR & RC & ACC & AUC & PR & RC   \\
  \hline
  \hline
  w/o DL & - & - & - & - & 71.6 & 89.7 & 74.7 & 75.2 \\
  \hline
  DenseNet &  57.4 &  68.4 &  55.9 & 56.2 & 75.8 & 87.4 & 74.7 & 76.0  \\
  ResNet18 & 48.9  & 66.6 &  46.3 & 48.0 & 73.7 & 89.3 & 76.1 & 76.7 \\
  AlexNet & 57.0 &  64.8 &  60.3 &  54.7 & 77.7 & \textbf{89.8} & 76.5 & 75.2  \\
  MobileNet & 57.0 &  64.4 &  57.7 &  54.6 & 71.6 & 89.3 & 74.7 & 75.2  \\
   \rowcolor[gray]{0.9} ViT  &  61.3 &  71.9 &  56.2 &  56.6 & \textbf{81.9} & 89.3 & \textbf{82.4} & \textbf{82.7} \\
  \hline
  \hline
  \end{tabular}
\label{tab:compare_radiomics}
\end{table}
Metrics for evaluating the models' clasification performance include Accuracy (ACC), the Area Under the receiver operating characteristic Curve (AUC), Precision (PR), and Recall (RC). Higher values in these metrics indicate better performance. We boost the classification results with the combination of radiomics and deep learning classifiers and obtain new SOTA results (Table 2). Despite the success of our proposed ensemble, we also identify some challenging cases where our classifiers fail to identify the existence and type of the IPMNs in Figure 3 (the second row for the failure cases compared to the first row for successfully predicted cases).

\begin{figure}
    \centering
    \includegraphics[width=0.9\textwidth]{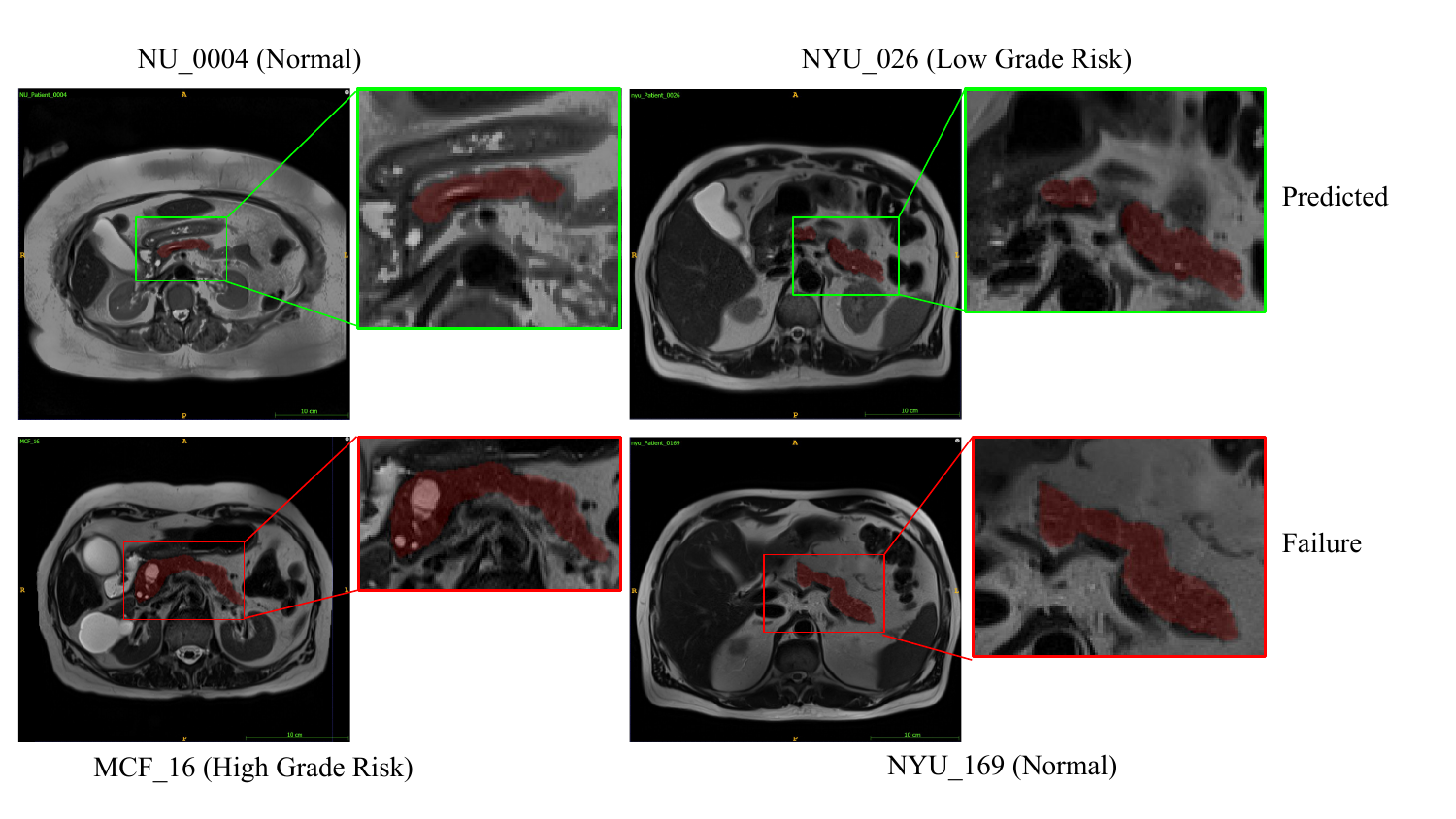}
    \caption{First and second rows show MRI cases where IPMN diagnosis and stratification are done correctly and incorrectly, respectively. Zoomed versions of the pancreas regions demonstrate the shape and appearance diversity.}
    \label{fig:goodbad}
    \vspace{-2mm}
\end{figure}

We assess the performance of a deep learning (DL) model and a radiomics model individually and in combination, for the classification of IPMN using multi-center multi-contrast MRI data. On a single center scenario, the DL model performs comparably to previous literature~\cite{salanitri2022neural}. However, when the models are tested on multi-center data, the performance of the DL model decreases, likely due to the heterogeneity of multi-center data compared to a single institution. We also observe that the decision fusion of the radiomics predictions to the DL model prediction improves its performance on multi-center data. This improvement can be attributed to the domain knowledge contained in the handcrafted radiomics features more than the deep features which are high-dimensional and sparse. Radiomics features are designed to capture important characteristics of IPMNs in dense vectors with more controlled variations, indicating a better generalization over multi-center data. The fusion method can be also viewed as a trainable linear layer on probability vectors. Lastly, our findings suggest that the information captured by the DL layers and radiomics features is complementary to a certain degree, and combining them can yield better performance than using either approach alone. 

To understand this further, we run experiments of radiomics classifiers with different combinations of features: T1 radiomics, T2 radiomics, combined T1+T2 radiomics, and T1+T2+clinical features. We achieve the following accuracies: 0.573, 0.650, 0.666, and 0.674, respectively, indicating that T1 and T2 features are complementary to each other, and the clinical feature (pancreas volumetry) increases the prediction performance.

\vspace{-2mm}
\section{Conclusion}
\vspace{-2mm}
IPMN cysts are a ticking time bomb that can progress into pancreatic cancer. Early detection and risk stratification of these precancerous lesions is crucial for effective treatment planning and disease control. However, this is no easy feat given the irregular shape, texture, and size of the cysts and the pancreas. To tackle this challenge, we propose a novel  CAD pipeline for IPMN risk classification from multi-contrast MRI scans. The proposed CAD system includes a self-adapting volumetric segmentation strategy for pancreas delineation and a newly designed deep learning-based classification scheme with a radiomics-based predictive approach at the decision level. In a series of rigorous experiments on multi-center data sets (246 MRI scans from five centers), we achieve unprecedented performance (81.9\% accuracy with radiomics) that surpasses the state of the art in the field (ViT 61.3\% without radiomics). Our ablation studies further underscore the pivotal role of both radiomics and deep learning modules for attaining superior performance compared to international guidelines and published studies, and highlight the importance of pancreas volume as a clinical feature.
\vspace{1cm}
\bibliographystyle{splncs04}
\bibliography{ref}
\end{document}